\pdfoutput=1

\documentclass[11pt,twoside,a4paper,cmspaper,final,collab]{cms-tdr}
\def\svnVersion{101809:102807}\def\cmsCernNo{2012-025}\def\cmsCernDate{\today}\def\cmsMessage{Submitted to Physical Review Letters}

\begin{document}\cmsNoteHeader{HIG-11-025}

\hyphenation{had-ron-i-za-tion}
\hyphenation{cal-or-i-me-ter}
\hyphenation{de-vices}

\RCS$Revision: 102807 $
\RCS$HeadURL: svn+ssh://alverson@svn.cern.ch/reps/tdr2/papers/HIG-11-025/trunk/HIG-11-025.tex $
\RCS$Id: HIG-11-025.tex 102807 2012-02-09 14:26:53Z alverson $

\newcommand{\Zee}{\ensuremath{\cPZ\to\Pep\Pem}}%\mbox{$\rm Z\rightarrow\rm e^{+} \rm e^{-}$}}
\newcommand{\nanob}{\mbox{{\rm ~nb}~}}
\newcommand{\fb}{\ensuremath{\mathrm{fb}}}
\newcommand{\pb}{\ensuremath{\mathrm{pb}}}
\newcommand{\ifb}{\ensuremath{\mathrm{fb^{-1}}}}
\newcommand{\ipb}{\ensuremath{\mathrm{pb^{-1}}}}
\newcommand{\Hmu}{\ensuremath{\PH\to\cPZ\cPZ^{(\ast)}\to 4\Pgm}}%{\ensuremath{\mathrm{H}\to\mathrm{ZZ^{(\ast)}}\to4\mu}}
\newcommand{\mmu}{\ensuremath{m_{4\Pgm}}}%{\ensuremath{\mathrm{m_{4\mu}}}}
\newcommand{\mH}{\ensuremath{m_{\PH}}}%{\ensuremath{m_{\mathrm{H}}}}
\newcommand{\lnQ}{\ensuremath{\mathrm{ln}(Q)}}
\newcommand{\mlnQ}{\ensuremath{-2\,\ln(Q)}}
\newcommand{\LS}{\ensuremath{\mathcal{L}(S)}}
\newcommand{\LB}{\ensuremath{\mathcal{L}(B)}}
\newcommand{\LSB}{\ensuremath{\mathcal{L}(S+B)}}
\newcommand{\mZ}{\ensuremath{m_{\cPZ}}}
\newcommand{\Zbb}{\ensuremath{(\cPZ/\Pgg^*){\cPqb\cPaqb}}}
\newcommand{\ZZ}{\ensuremath{\cPZ/\Pgg^*\cPZ/\Pgg^*}}
\newcommand{\Zgam}{\ensuremath{\cPZ/\Pgg^*}}
\newcommand{\mumu}{\ensuremath{\Pgmp\Pgmm}}
\cmsNoteHeader{HIG-11-025} % This is over-written in the CMS environment: useful as preprint no. for export versions
\title{Search for the standard model Higgs boson in the decay channel \texorpdfstring{$\PH\rightarrow\cPZ\cPZ\rightarrow 4\ell$ in $\Pp\Pp$ collisions at $\sqrt{s} = 7\TeV$}{H to ZZ to 4 leptons in pp collisions at sqrt(s) = 7 TeV}}

\date{\today}

\abstract{A search for a Higgs boson in the four-lepton decay channel $\PH \rightarrow \cPZ\cPZ$,
with each \cPZ\ boson decaying to an electron or muon pair, is reported.
The search covers Higgs boson mass hypotheses in the range $110 < \mH < 600\GeV$.
The analysis uses data corresponding to an integrated luminosity of $4.7\fbinv$ recorded by the CMS detector
in pp collisions at $\sqrt{s}$ = 7\TeV from the LHC.
Seventy-two events are observed with four-lepton invariant mass $m_{4\ell} > 100\GeV$
(with thirteen below $160\GeV$), while $67.1 \pm 6.0$ ($9.5 \pm 1.3$) events are expected from
background.
The four-lepton mass distribution is consistent with the expectation of standard model background
production of \cPZ\cPZ\ pairs.
Upper limits at 95\% confidence level exclude the standard model Higgs boson in the ranges
134--158\GeV, 180--305\GeV, and 340--465\GeV.
Small excesses of events are observed around masses
of 119, 126, and 320\GeV, making the observed limits
weaker than expected in the absence of a signal.}

\hypersetup{%
pdfauthor={CMS Collaboration},%
pdftitle={Search for the standard model Higgs boson in the decay channel H to ZZ to 4l in pp collisions at sqrt(s) = 7 TeV}, %
pdfsubject={CMS},%
pdfkeywords={CMS, physics, Higgs, multilepton, diboson}}

\maketitle %maketitle comes after all the front information has been supplied

The standard model (SM) of electroweak
interactions~\cite{StandardModel67_1,StandardModel67_2,StandardModel67_3}
relies on a scalar particle,  the Higgs boson, associated with the field responsible for the
spontaneous electroweak symmetry
breaking~\cite{Englert:1964et,Higgs:1964ia,Higgs:1964pj,Guralnik:1964eu,Higgs:1966ev,Kibble:1967sv}.
The existence of the Higgs boson has yet to be established
experimentally, while its mass, $\mH$, is not fixed by the theory.
Direct searches for the SM Higgs boson at the LEP $\Pep\Pem$ collider and the
Tevatron $\Pp\Pap$ collider have led, respectively, to a lower mass bound of
$\mH > 114.4\GeV$~\cite{Barate:2003sz}, and to an exclusion in the
range 162--166\GeV~\cite{TEVHIGGS_2010}, at 95\% CL.
Indirect constraints from precision measurements favour the mass range
$\mH < 158\GeV$~\cite{:2010vi} at 95\% CL.
The inclusive Higgs boson production followed by the decay $\PH\rightarrow\cPZ\cPZ$ is expected to be one
of the main discovery channels at the CERN proton-proton (\Pp\Pp) Large Hadron Collider (LHC)
for a wide range of $\mH$ values.
Using the $\PH\rightarrow\cPZ\cPZ$ and the $\PH\rightarrow\PW\PW$ decay channels,
the ATLAS collaboration have excluded at 95\% CL the mass ranges
145--206\GeV, 214--224\GeV, and 340--450\GeV~\cite{Aad:2011uq,ATLAS:2011af,ATLAS:2011aa}.

In this Letter, an inclusive search in the four-lepton decay channel,
$\PH\rightarrow\cPZ\cPZ\rightarrow  \ell^{+}\ell^{-}\ell'^{+}\ell'^{-}$
with $\ell, \ell' = \Pe$ or $\Pgm$, abbreviated as $\PH\rightarrow 4\ell$, is presented.
The analysis is designed for a Higgs boson in the mass range $110 < \mH < 600\GeV$.
It uses pp data from the LHC collected at $\sqrt{s}$ = 7\TeV
by the Compact Muon Solenoid (CMS) experiment during 2010 and 2011.
The data corresponds to an integrated luminosity of $4.7\fbinv$.
The search relies solely on the measurement of leptons, and the analysis achieves high lepton reconstruction,
identification, and isolation efficiencies for a $\cPZ\cPZ\rightarrow 4\ell$ system composed of two pairs of
same-flavour and opposite-charge isolated leptons, $\Pep\Pem$ or $\Pgmp\Pgmm$, in the measurement range
$m_{4\ell} > 100\GeV$. One or both of the \cPZ\ bosons can be off-shell.
The background sources include an irreducible four-lepton contribution from direct \cPZ\cPZ\ production
via $\Pq\Paq$ annihilation and $\Pg\Pg$ fusion.
Reducible contributions arise from $\cPZ\cPqb\cPaqb$ and $\ttbar$
where the final states contain two isolated leptons and two $\cPqb$ jets producing secondary leptons.
Additional background of instrumental nature arises from $\cPZ+\text{jets}$ events where jets are misidentified
as leptons.

Particles produced in the pp collisions are detected in the pseudorapidity range $ | \eta | < 5$,
where $\eta = - \ln \tan (\theta / 2) $ and $\theta$ is the polar angle with respect to the direction
of the proton beam.
The CMS detector comprises a superconducting solenoid, providing a uniform magnetic
field of 3.8\unit{T} in the bore, equipped with silicon pixel and strip tracking systems  ($ | \eta | < 2.5$)
surrounded by a lead tungstate crystal electromagnetic  calorimeter (ECAL) and a brass-scintillator
hadronic calorimeter (HCAL) ($ | \eta | < 3.0$).
A steel/quartz-fiber Cherenkov calorimeter extends the coverage ($ | \eta | < 5$).
The steel return yoke outside the solenoid is instrumented with gas detectors used to
identify muons ($ | \eta | < 2.4$).
A detailed description of the detector is given in Ref.~\cite{:2008zzk}.

Monte Carlo (MC) samples for the SM Higgs boson signal and for background processes are used
to optimize the event selection and to evaluate the acceptance and systematic uncertainties.
The Higgs boson signals from gluon-fusion ($\Pg\Pg \rightarrow \PH$),
and vector-boson fusion ($\Pq\Paq \rightarrow \Pq\Paq \PH$), are generated with
\POWHEG~\cite{powheg} at next-to-leading-order (NLO) and a dedicated
generator from Ref.~\cite{Gao:2010qx}.
Additional samples of \PW\PH, \cPZ\PH, and $\ttbar\PH$ events are generated with
\PYTHIA~\cite{Sjostrand:2006za}.
Events at generator level are reweighted according to the total cross section $\sigma(\Pp\Pp\rightarrow \PH)$,
which contains contributions from gluon fusion up to next-to-next-to-leading-order (NNLO) and next-next-to-leading-log taken from
Refs.~\cite{Anastasiou:2008tj,deFlorian:2009hc,LHCHiggsCrossSectionWorkingGroup:2011ti,Dawson:1990zj,Spira:1995rr,Harlander:2002wh,Anastasiou:2002yz,Ravindran:2003um,Catani:2003zt,Actis:2008ug}
and from the  weak-boson fusion contribution computed at NNLO in
Refs.~\cite{LHCHiggsCrossSectionWorkingGroup:2011ti,Ciccolini:2007jr,Ciccolini:2007ec,Figy:2003nv,Arnold:2008rz,Bolzoni:2010xr}.
The total cross section is scaled by the branching fraction
$\mathcal{B}(\PH\rightarrow 4\ell)$~\cite{LHCHiggsCrossSectionWorkingGroup:2011ti,Bredenstein:2006rh,Bredenstein:2006ha,hdecay2,Actis:2008ts}.
Interference effects in the $4\Pe$ and $4\Pgm$ channels are calculated with
\textsc{prophecy4f}~\cite{LHCHiggsCrossSectionWorkingGroup:2011ti,Bredenstein:2006rh,Bredenstein:2006ha}.
The SM background contribution from \cPZ\cPZ\ production via $\Pq\Paq$ is generated at NLO with \POWHEG, while
other di-boson processes (\PW\PW, \PW\cPZ) are generated with \PYTHIA
with cross sections rescaled to NLO predictions.
The $\Pg\Pg \rightarrow\cPZ\cPZ$ contribution is generated with \textsc{gg2zz}~\cite{Binoth:2008pr}.
The $\cPZ\cPqb\cPaqb$, $\cPZ\cPqc\cPaqc$, $\cPZ\Pgg$, and $\cPZ+\text{light jets}$ samples
are generated with \MADGRAPH~\cite{Alwall:2007st}
with cross sections rescaled to NNLO prediction for inclusive \cPZ\ production.
The $\ttbar$ events are generated at NLO with \POWHEG.
The generation takes into account the internal initial state
and final state radiation (FSR) effects which can lead to the presence of additional
hard photons in an event.
All events are processed through a detailed simulation of the CMS detector based on
\GEANTfour~\cite{GEANT} and are reconstructed with the same algorithms that are used for data.

Collision events are selected by the trigger system that requires the presence of a
pair of electrons (a pair of muons) with transverse energy (transverse momenta) for
the first and second lepton above  17 and 8\GeV respectively.
The trigger efficiency within the acceptance of this analysis is greater than 99\% for signal in the 4\Pe\ and
4\Pgm\ channels, and rises from about 97.5\% at $\mH = 120\GeV$ to
above 99\% at $\mH > 140\GeV$ in the $2\Pe2\Pgm$ channel, within the
acceptance of this analysis.

Electrons are reconstructed within the geometrical acceptance, $|\eta^{\Pe}| < 2.5$, and with $\PT^{\Pe} > 7\GeV$,
by combining information from the ECAL and inner tracker~\cite{Baffioni:2006cd,CMS-PAS-EGM-10-004}.
Electron identification selection requirements rely on electromagnetic shower-shape observables
and on observables combining tracker and calorimetry information.
The selection criteria depend on $\PT^{\Pe}$ and $| \eta^{\Pe} |$, and on a categorization
according to observables sensitive to the amount of bremsstrahlung emitted along the trajectory
in the inner tracker.
Muons are reconstructed~\cite{CMS-PAS-MUO-10-002} within $|\eta^{\Pgm}| < 2.4$ and $\PT^{\Pgm} > 5\GeV$,
using information from both the inner tracker and the muon spectrometer.
The inner track is required to be composed of more than 10 tracker-layer hits
to ensure a precise  measurement of the momentum.
The efficiencies are measured in data, using a tag-and-probe technique~\cite{CMS:2011aa}
based on an inclusive sample of \cPZ\ events.
The measurements are performed in several ranges in $\PT^{\ell} $ and $ |\eta| $.
The product of reconstruction and identification efficiencies for electrons in the ECAL barrel (endcaps)
vary from about  68\% (62\%) for $7 < \PT^{\Pe} < 10\GeV$ to  82\% (74\%) at $\PT^{\Pe} \simeq 10\GeV$,
and reach 90\% (89\%) for $\PT^{\Pe} \simeq 20\GeV$. It drops to about 85\% in the transition region,
$1.44 < |\eta| < 1.57$, between the ECAL barrel and endcaps.
The muons are reconstructed and identified with efficiencies above ${\sim}98\%$.
Lepton candidates are defined with a loose constraint on their isolation, by requiring the sum of the transverse
momenta of tracks $i$ within a cone around the lepton $\ell$ of $\Delta R = \sqrt{(\eta^\ell - \eta^i)^{2} + (\phi^\ell- \phi^i)^{2}} < 0.3$,
where $\phi$ is the azimuthal angle, to have $\sum_{i}  p_{T,{\rm track}}^i /  p_T^{\ell}  < 0.7$.
The lepton isolation efficiency for identified leptons with this very loose definition of isolation
is found to be greater than $99\%$.

We first require a \cPZ\ candidate formed with a pair of lepton
candidates satisfying $50 < m_{1,2} < 120\GeV$, $\PT^{\ell_1} > 20\GeV$, and
$\PT^{\ell_2} > 10\GeV$.
The \PT thresholds ensure that the leptons are on the high-efficiency plateau for the trigger.
The lepton pair is required to be well isolated using a combination of the tracker, ECAL and
HCAL information.
The sum of the combined relative isolation $R_\text{iso}$ for the two leptons is required to satisfy
$R_\text{iso}^{\ell_1} +  R_\text{iso}^{\ell_2} < 0.35$, where for each lepton,
$R_\text{iso} = (1/ \PT^{\ell}) \times \left( \sum_{i}  p_{\mathrm{T},\text{track}}^i + \sum_{j}  E_{\mathrm{T},\text{ECAL}}^j
                                                                 + \sum_{k}  E_{\mathrm{T},\text{HCAL}}^k \right)$,
with sums running over the charged tracks $i$, and the \ET from energy deposits in cells
$j$ and $k$ of the ECAL and HCAL within a cone of radius  $\Delta R  < 0.3$, respectively.
The footprint of the lepton is removed from the isolation sum.
The combined isolation efficiencies measured with data using the tag-and-probe technique
are found to be $>99\%$ for muons and between 94\% to 99\% for electrons.
The isolation is made largely insensitive to the number of overlapping
\Pp\Pp\ interactions by correcting for the average energy flow~\cite{Cacciari:2008gn} per unit area
measured as a function of the number of primary vertices.
The ratio of the efficiencies measured with data and with simulated
$\cPZ\rightarrow \ell \ell$ events is found to be consistent with unity.
The significance of the signed impact parameter of each lepton relative to the event
vertex, ${\text{SIP}_\text{3D}} = \frac{\text{IP}}{\sigma_\text{IP}}$,
where IP  is the impact parameter in three dimensions and
$\sigma_\text{IP}$ the associated uncertainty, is required to satisfy
$| {\text{SIP}_\text{3D}} | < 4$.
The $\ell^+ \ell^-$ pair with reconstructed mass closest to the nominal \cPZ\ boson mass
is retained and denoted ${\cPZ_1}$.
The $\cPZ_1$+X dataset thus defined contains the samples used to estimate reducible and
instrumental backgrounds, as well as the \cPZ\cPZ\ rates.
In the next step, a subset of events is identified with at least a third lepton candidate.
The ${\cPZ_1} + \ell$ events  are used to measure misidentified lepton rates.
A subset of events with at least a fourth lepton candidate of any flavour or charge is then identified.
Together, the ${\cPZ_1} + \ell$ and ${\cPZ_1} + \ell \ell'$ samples provide ways to estimate the remaining
reducible ($\cPZ\bbbar$, $\ttbar$) and instrumental ($\cPZ+\text{light jets}$) backgrounds.
For the signal, we select a second lepton pair, denoted ${\cPZ_2}$, from the remaining
same flavour $\ell^+ \ell^-$ combinations, by requiring $m_{\cPZ_2} > 12\GeV$,
with the restriction $m_{4\ell} > 100\GeV$.
For the $4\Pe$ and $4\Pgm$ final states, at least three of the four combinations of opposite-sign
pairs must satisfy $m_{\ell\ell} > 12\GeV$.
If more than one ${\cPZ_2}$ candidate satisfies all criteria, the ambiguity
is resolved by choosing the leptons of highest \PT.
The isolation and impact parameter are used to further suppress the remaining backgrounds.
We require for any combination of two leptons $i$ and $j$, irrespective of flavour or charge, that
$R_{\text{iso},i} +  R_{\text{iso},j} < 0.35$ and also impose $| {\text{SIP}_\text{3D}} | < 4$ for each of the four leptons.

Finally, to select the four-lepton signal candidates, we require that
the ${\cPZ_1}$ and  ${\cPZ_2}$ masses satisfy
$m_{\cPZ_1}^{\rm min} < m_{\cPZ_1} < 120\GeV$ and $m_{\cPZ_2}^\text{min} < m_{\cPZ_2} < 120\GeV$,
with $(m_{\cPZ_1}^\text{min}, m_{\cPZ_2}^\text{min})  = (50,12)\GeV$ defining the  baseline selection and
$(m_{\cPZ_1}^\text{min}, m_{\cPZ_2}^\text{min}) = (60,60)\GeV$ defining the  high-mass selection.
The baseline selection is used to search for the Higgs boson, and the high-mass selection is used to
measure the \cPZ\cPZ\ cross section.

The event yields are found to be in good agreement with the MC background expectation at each step
of event selection. The \cPZ\cPZ\ and \cPZ+X backgrounds dominate after the full event selection.
The overall signal detection efficiency for the $4\Pe$ ($4\Pgm$, $2\Pe2\Pgm$) channel is evaluated
by MC and increases
from ${\approx}21\%$~(59\%, 35\%) at $\mH = 120\GeV$
to  ${\approx}35\%$~(71\%, 50\%) at $\mH = 140\GeV$,
reaching a plateau at ${\approx}51\%$~(81\%, 63\%) at $\mH = 200\GeV$,
and then slowly rising to ${\approx}60\%$~(83\%, 72\%) at $\mH = 350\GeV$.
The relative mass resolution estimated from MC signal samples is
about 2.1\% (1.1\%, 1.6\%) for $4\Pe$~($4\Pgm$, $2\Pe2\Pgm$).

The small number of observed events precludes a precise direct evaluation of background
by extrapolating from mass sidebands.
Instead, we rely on MC to evaluate the number of events expected from the \cPZ\cPZ\ background.
The cross section for \cPZ\cPZ\ production at NLO, through the dominant process of $\Pq\Paq$ annihilation
and through $\Pg\Pg$ fusion, is calculated with \textsc{mcfm}~\cite{MCFM,Campbell:1999ah,Campbell:2011bn}.
The theoretical uncertainties are computed as a function of $m_{4\ell}$, varying both the QCD renormalization
and factorization scales and the parton distribution functions (PDF) set following the PDF4LHC
recommendations~\cite{Botje:2011sn,Alekhin:2011sk,Lai:2010vv,Martin:2009iq,Ball:2011mu}.
The uncertainties for the QCD and PDF scales for each final state are on average 8\%.
The number of predicted $\cPZ\cPZ\rightarrow 4\ell$  events and their uncertainties after the baseline
selection are given in Table~\ref{tab:FinalYields}.
As a consistency check, an evaluation is made based on a normalization to the measured
inclusive single-\cPZ\ production, a procedure discussed in
Refs.~\cite{Baffioni:2006sk,CMS_NOTE_2006-122}.
The measured rate of single Z bosons defined in this analysis is used to predict the total \cPZ\cPZ\ rate;
making use of the ratio of the theoretical cross sections for \cPZ\cPZ\ and \cPZ\ production,
and the ratio of the reconstruction and selection efficiencies for the four-lepton  and two-lepton final states.
The results are in agreement with the \cPZ\cPZ\ rates reported in Table~\ref{tab:FinalYields} within uncertainties.

To estimate the reducible ($\cPZ\bbbar$, $\ttbar$) and instrumental ($\cPZ+\text{light jets}$)
backgrounds, a region well separated from the signal region is defined
by relaxing and inverting some selection criteria and verifying that the event rates change
according to MC expectation.
The event rates measured in the background control region are then extrapolated to the
signal region.
The control region for \cPZ+X, where X stands for $\bbbar$, $\cPqc\cPaqc$, gluon or light quark jets,
is obtained by relaxing the isolation and identification criteria for two additional reconstructed lepton
objects (a measured track for muons, or a combination of a track and a cluster of ECAL energy
deposits  for electrons) indicated as $\ell_\text{reco}\ell_\text{reco}$.
The additional pair of leptons must have like sign charge (to avoid signal contamination) and same flavour
($\Pe^{\pm}\Pe^{\pm}, \Pgm^{\pm}\Pgm^{\pm}$), a reconstructed invariant mass $m_{\cPZ_2}$
either satisfying the baseline selection or the high-mass selection, and $m_{4\ell} > 100\GeV$.
A sample $\cPZ_1 + \ell_\text{reco}$, with at least one reconstructed lepton object, is also defined
for the measurement of the lepton misidentification probability,  the probability for a
reconstructed object to pass the isolation and identification requirements.
The contamination from \PW\cPZ\ in this set of events is suppressed by requiring that the imbalance
of the measured energy deposition in the transverse plane is below $25\GeV$.
From the $\cPZ+\ell_\text{reco} \ell_\text{reco}$ sample the expected number of \cPZ+X background events in the
signal region is obtained by taking into account the lepton misidentification probability for each
of the two additional leptons.
The number of background events expected in the signal region, normalized to
the integrated luminosity, and the associated systematic uncertainties, are given
in Table~\ref{tab:FinalYields} for the baseline selection in the range
$100 < m_{4\ell} < 600\GeV$.
The reducible and instrumental background is found to be dominated by $\cPZ+\text{light jets}$.
A small residual contamination of $\cPZ\bbbar$ remains at low mass while for the high-mass
selection these reducible backgrounds are an order of magnitude smaller and therefore can be neglected.
This was verified by performing a measurement of $\cPZ\bbbar$ and $\ttbar$ rates
in a dedicated four-lepton background control region, defined by requiring a ${\cPZ_1}$ and two additional leptons
satisfying an inverted ${\text{SIP}_\text{3D}} $ requirement, namely $ \abs{ {\text{SIP}_\text{3D}} } > 5$, and with
relaxed isolation, charge, and flavour requirements.
This ensures a negligible $\cPZ+\text{light jets}$ contribution in the four-lepton background control region, while
the signal and the \cPZ\cPZ\ background are absent.
To extract background rates, the reconstructed $\cPZ_1$  mass for the sum of the $\cPZ_1+2\Pe$,
$\cPZ_1+2\Pgm$, and $\cPZ_1+\Pe\Pgm$  final states is fit with a Breit-Wigner function
convoluted with a Crystal Ball function~\cite{refCrystalBall} for the $\cPZ_1$ peak from $\cPZ\bbbar$
and Chebychev polynomials for the description of the $\ttbar$ continuum.
The extrapolation to the signal region relies on knowledge of, and the distinct features of, the $\text{SIP}_\text{3D}$
distributions for the ${\cPZ_2}$ leptons of the $\ttbar$  and $\cPZ\bbbar$ backgrounds.
The result is found to be compatible with the MC expectation in the signal region within the
systematic uncertainty of 20\%.

Systematic uncertainties are evaluated from data for trigger efficiency (1.5\%), lepton reconstruction
and identification (2\,--\,3\%), and isolation efficiencies (2\%).
Systematic uncertainties on energy-momentum calibration (0.5\%), and energy resolution
are accounted for by their effect on the reconstructed mass distributions. The effect of the
energy resolution uncertainties is taken into account by introducing a 30\% uncertainty
on the width of the signal mass peak.
Additional systematic uncertainties arise from limited statistics in the reducible
background control regions.
All reducible and instrumental background sources are derived from control regions, and the comparison
of data with the background expectation in the signal region is independent of the
uncertainty on the LHC integrated luminosity of the data sample.
This uncertainty (4.5\%)~\cite{lumiPAS} enters the evaluation of the \cPZ\cPZ\ background
and in the calculation of the cross section limit through the normalization of the signal.
Systematic uncertainties on the Higgs boson cross section (17\,--\,20\%) and
branching fraction (2\%) are taken from Ref.~\cite{LHCHiggsCrossSectionWorkingGroup:2011ti}.

Recent studies \cite{Passarino:2010qk,LHCHiggsCrossSectionWorkingGroup:2011ti,Anastasiou:2011pi}
show that current Monte Carlo simulations do not describe the expected Higgs boson mass line shape
above ${\approx}300\GeV$.
These effects are estimated to amount to an additional uncertainty on the theoretical cross section,
and hence on the limits, of about 4\% at $\mH = 300\GeV$ and 10\,--\,30\% for
$\mH$ of 400--600$\GeV$.

\begin{figure}[tbhp]
 \begin{center}
\includegraphics[width=0.49\textwidth]{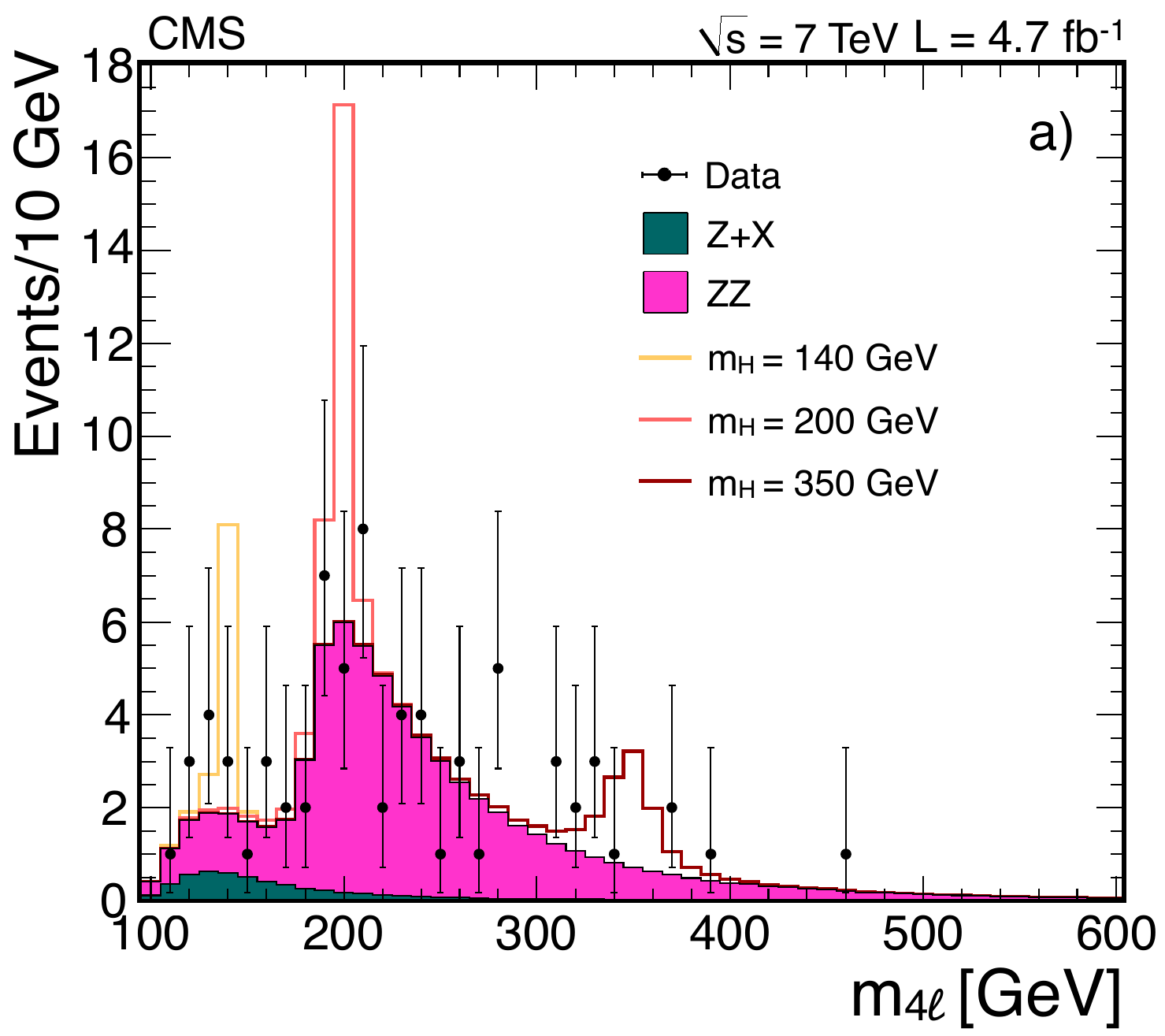}
\includegraphics[width=0.49\textwidth]{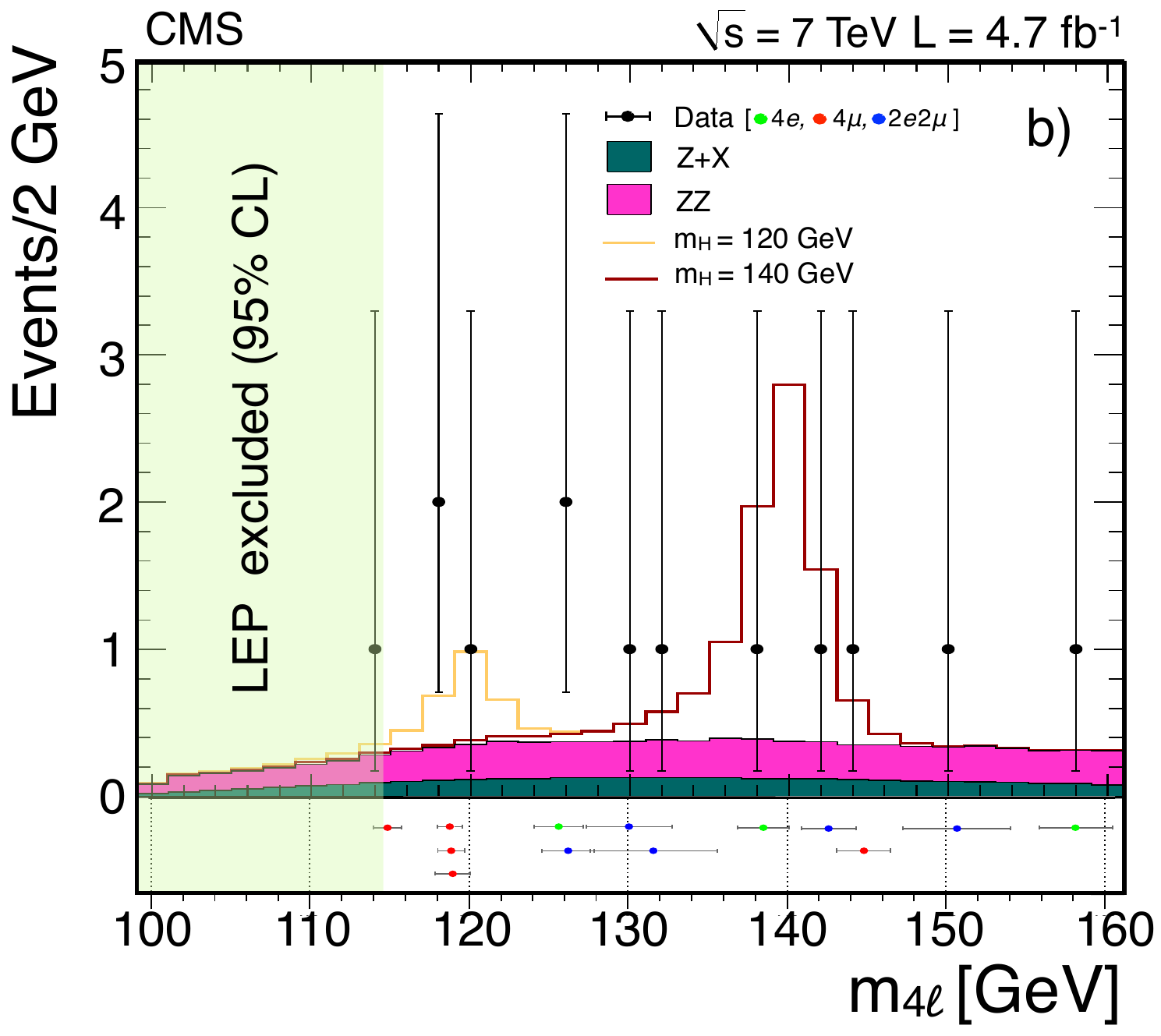}
   \caption{
    a) Distribution of the four-lepton reconstructed mass for the sum of the
          $4\Pe$, $4\Pgm$, and $2\Pe2\Pgm$ channels.
    b) Expansion of the low mass range with existing exclusion limits at 95\% CL; also shown are the central
          values and individual candidate mass measurement uncertainties.
          Points represent the data, shaded histograms represent the background and unshaded histogram the
          signal expectations.}
   \label{fig:Mass4lBaseline}
 \end{center}
\end{figure}

\begin{table}[b]
\begin{center}
\caption{The number of candidates observed, compared to background and signal rates for each final state
                for  $100 < m_{4\ell} < 600\GeV$ for the baseline selection.
                For the Z+X background, the estimations are based on data
                }
\label{tab:FinalYields}
\begin{tabular}{l|c|c|c}
\hline
\textbf{Channel} & $4\Pe$ & $4\Pgm$ & $2\Pe2\Pgm$  \\
\hline
ZZ background &   12.27 $\pm$ 1.16 & 19.11 $\pm$ 1.75  & 30.25 $\pm$ 2.78                                 \\ % 4.7/fb
\cPZ+X & 1.67 $\pm$ 0.55 &  1.13 $\pm$ 0.55 &  2.71 $\pm$ 0.96                                      \\ % 4.7/fb
\hline
All background       & $13.94 \pm 1.28$ & $ 20.24 \pm 1.83$ & $32.96 \pm 2.94$       \\ % 4.7/fb
\hline
$\mH = 120\GeV$ & 0.25 & 0.62 & 0.68  \\% 4.7/fb
$\mH = 140\GeV$ & 1.32 & 2.48 & 3.37 \\% 4.7/fb
$\mH = 350\GeV$ & 1.95 & 2.61 & 4.64 \\ % 4.7/fb
\hline
Observed & 12 & 23  & 37\\ % 4.7/fb
\hline
\end{tabular}
\end{center}
\end{table}
The number of candidates observed, as well as the estimated background in the signal region,
are reported in Table~\ref{tab:FinalYields} for the baseline selection.
The reconstructed four-lepton invariant mass distribution for the combined $4\Pe$, $4\Pgm$, and  $2\Pe2\Pgm$ channels
with the baseline selection is shown in Fig.~\ref{fig:Mass4lBaseline}a and compared to expectations
from the backgrounds.
The shape of the mass distribution below $\mH = 180\GeV$ reflects the shape of the dominant
$\Pq\Paq$ annihilation process~\cite{Dittmaier:2012vm}.
The low mass range is shown in Fig.~\ref{fig:Mass4lBaseline}b together with the mass of each
candidate and its uncertainty.
The reducible and instrumental background rates are small.
These rates have been obtained from data and the corresponding $m_{4\ell}$
distributions are obtained from MC samples.

\begin{figure}[hbtp]
 \begin{center}
  \includegraphics[width=0.49\textwidth]{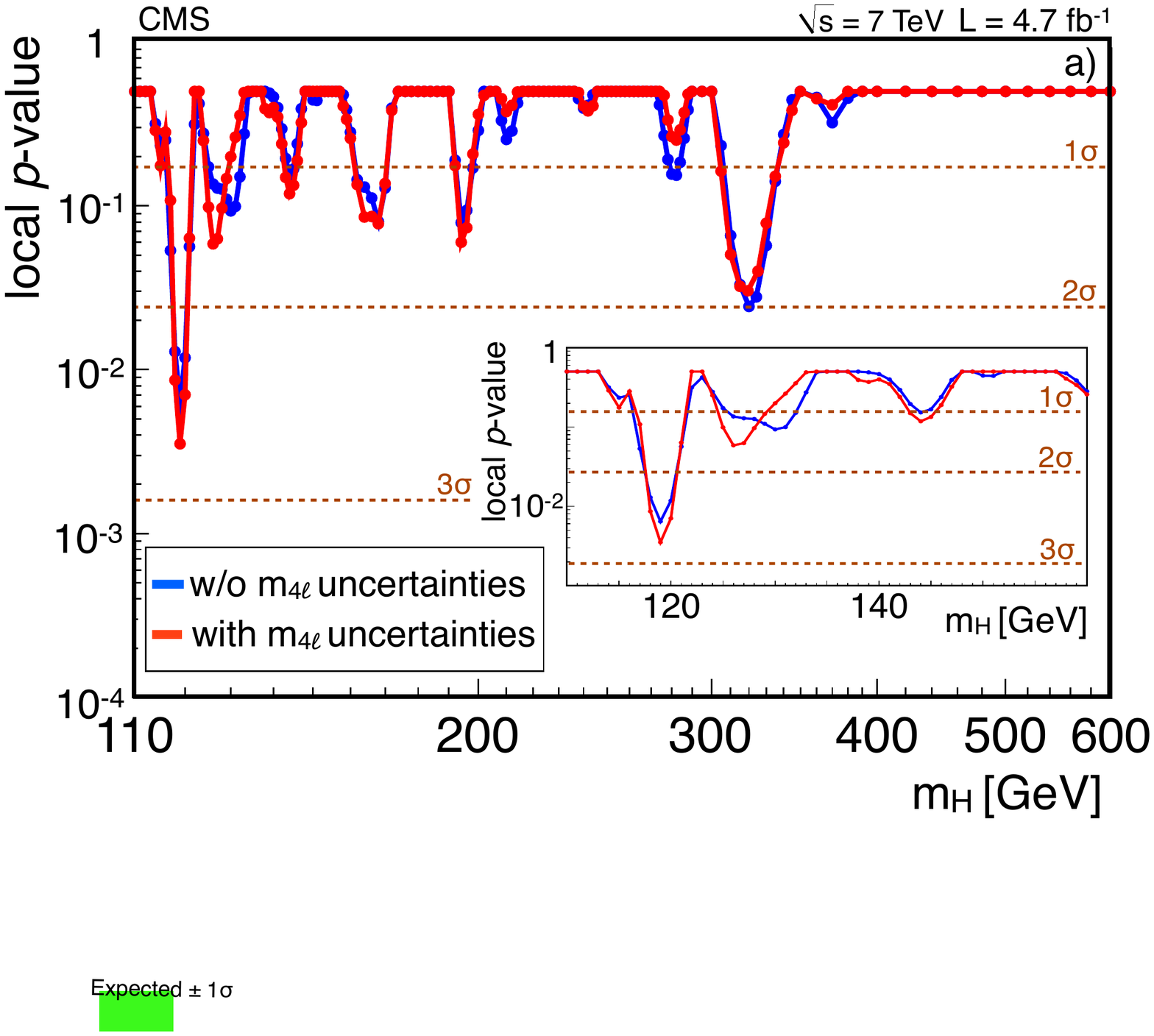}
  \includegraphics[width=0.49\textwidth]{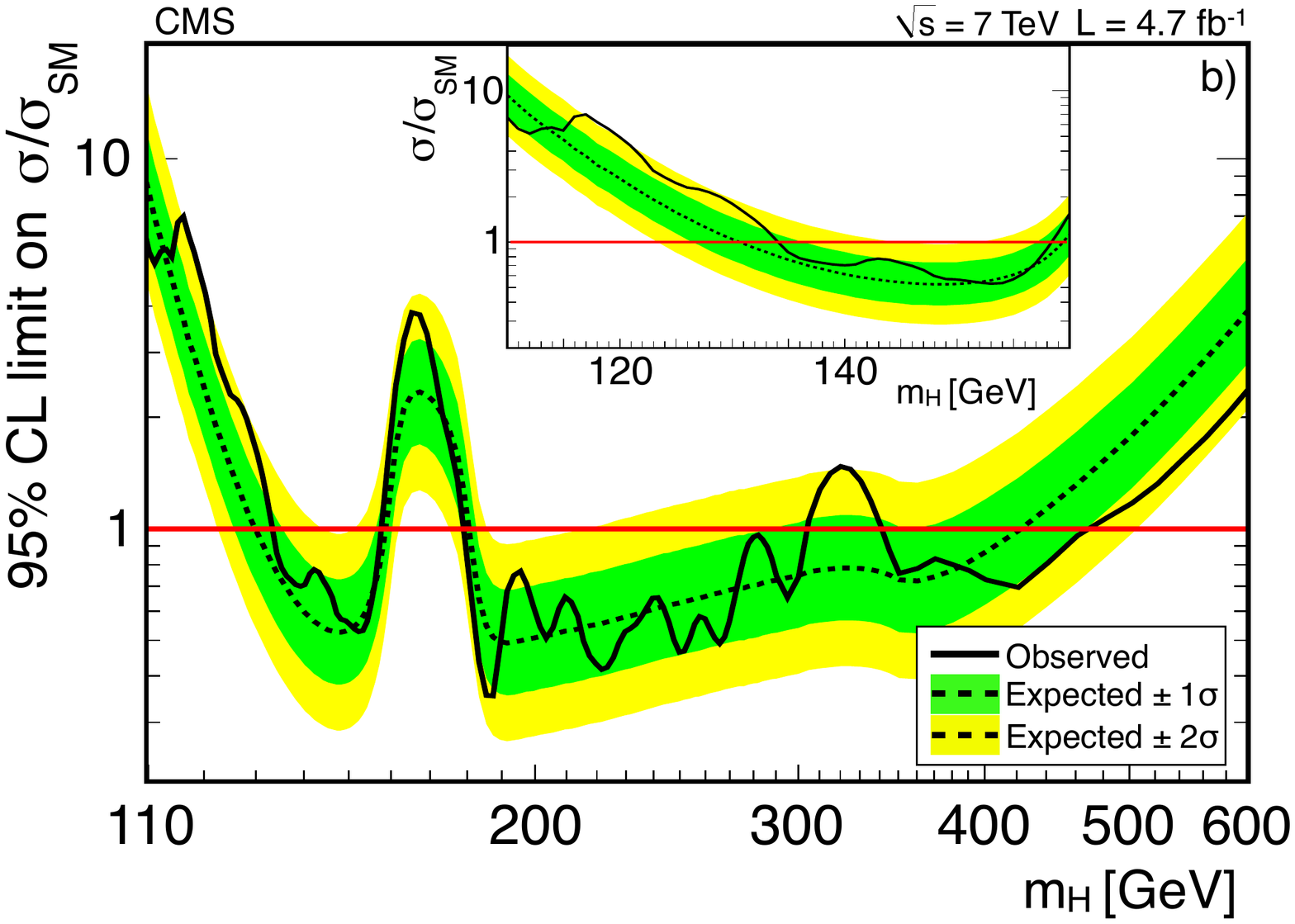}
   \caption{
         a)  The significance of the local excesses with respect to the standard model expectation as
          a function of the Higgs boson mass, without (blue) or with (red) individual candidate mass
          measurement uncertainties.
         b)  The observed and the median expected upper limits at 95\% CL  on
          $\sigma (\Pp\Pp \rightarrow \PH + X) \times \mathcal{B}(\cPZ\cPZ \rightarrow 4\ell)$,
          normalized to the standard model cross section values $\sigma_\text{SM}$,
          for a Higgs boson in the mass range 110--600$\GeV$, using the CL$_\mathrm{s}$ approach.
          The insets expand the low mass range.}
   \label{fig:xsLimCutandCountDATA}
 \end{center}
\end{figure}

The measured distribution is compatible with the expectation from SM direct
production of \cPZ\cPZ\ pairs.
We observe 72 candidates, 12 in $4\Pe$, 23 in $4\Pgm$, and 37 in $2\Pe2\Pgm$,
while $67.1 \pm 6.0$ events are expected from standard model background processes.
No hard photon ($\PT^{\Pgg} > 5\GeV$) was found, outside the isolation
veto cone that surrounds each lepton, that could be unambiguously identified
as FSR.
Thirteen candidates are observed within $100 < m_{4\ell} < 160\GeV$ while $9.5 \pm 1.3$
background events are expected.
We observe 53 candidates for the high-mass selection compared to an expectation of
$51.3 \pm 4.6$ events from background.
This high-mass event selection is used to provide a measurement of the total cross section
$ \sigma (\Pp\Pp \rightarrow {\cPZ\cPZ} + X) \times \mathcal{B}({\cPZ\cPZ} \rightarrow 4\ell)
      = 28.1  ^{+4.6 } _{-4.0}  (\text{stat.}) \pm 1.2 (\text{syst.}) \pm 1.3  (\text{lumi.})\unit{fb}$.
The measurement agrees with the SM prediction at NLO~\cite{MCFM} of $27.9 \pm1.9\unit{fb}$.
The local $p$-values, representing the significance of local excesses relative to the standard model
expectation, are shown as a function of $\mH$ in Fig.~\ref{fig:xsLimCutandCountDATA}a,
obtained  either taking into account or not the individual candidate mass measurement uncertainties,
for the combination of the three channels.
Excesses are observed for masses near $119\GeV$ and $320\GeV$.
The small ${\approx}2\sigma$ excess near $320\GeV$ includes three events with
$\PT^{4\ell} > 50\GeV$.
The most significant excess near $119\GeV$ corresponds to about $2.5\sigma$ significance.
The significance is less than $1.0\sigma$ (about  $1.6\sigma$) when the look-elsewhere
effect~\cite{Gross:2010qma} is accounted for over the full mass range
(for the low-mass range $100 < m_{4\ell} < 160\GeV$).
The local significances change only slightly when including candidate mass uncertainties,
instead of using the average mass resolution, e.g. rising to $2.7\sigma$ around $119\GeV$
and reaching $1.5\sigma$ around $126\GeV$.

In absence of a significant clustering of candidates at any given mass, we derive exclusion limits.
The exclusion limits for a SM-like Higgs boson are computed for a large number of mass points
in the mass range 110--600\GeV, using the predicted signal and background mass
distribution shapes.
The choice of the step size in the scan between Higgs mass hypotheses is driven
by either detector resolution, or the natural width of the Higgs boson.
The signal mass distributions shapes are determined using simulated samples
for 27 values of $\mH$ covering the full mass range.
The shapes are fit using a function obtained from a convolution of a Breit-Wigner probability density function
to describe the theoretical resonance line shape and a Crystal Ball function to account for the detector effects.
The parameters of the Crystal Ball function are interpolated for the $\mH$ points  where there is no
simulated sample available.
The shapes of the background mass distributions are determined by fits to the simulated sample of events,
while the normalization is taken from estimates of overall event yields as described above.
For each mass hypothesis, we perform an unbinned likelihood fit using the statistical
approach discussed in Ref.~\cite{ATL-PHYS-PUB-2011-011}.
We account for systematic uncertainties in the form of nuisance parameters with a log-normal
probability density function.
The observed and median expected upper limits on
$\sigma (\Pp\Pp \rightarrow \PH + X) \times \mathcal{B}(\PH \rightarrow \cPZ\cPZ)  \times \mathcal{B}(\cPZ\cPZ \rightarrow 4\ell)$
at 95\% CL are shown in Fig.~\ref{fig:xsLimCutandCountDATA}b.
The limits are calculated relative to the expected SM Higgs boson cross section values $\sigma_\text{SM}$, using the modified frequentist
method CL$_\mathrm{s}$~\cite{Junk:1999kv,Read:451614}.
The bands represent the $1\sigma$ and $2\sigma$ probability intervals around the expected limit.
These upper limits exclude the standard model Higgs boson at 95\% CL in the $\mH$
ranges 134--158\GeV, 180--305\GeV and 340--465\GeV.
The limits reflect the dependence of the branching ratio $\mathcal{B}(\PH \rightarrow \cPZ\cPZ)$
on $\mH$.
The worsening of the limits at high mass arises from the decreasing cross section for
the $\PH\rightarrow 4\ell$ signal.
By virtue of the excellent mass resolution and low background, the structure in the measured limits
follows the fluctuations of the number of observed events.

In summary, a search for the standard model Higgs boson has been presented in the four-lepton decay modes.
Upper limits at 95\% confidence level exclude the Higgs boson mass ranges
134--158\GeV, 180--305\GeV, and 340--465\GeV.
A major fraction of the explored mass range is thus excluded at 95\% CL and the
exclusion limits extend beyond the sensitivity of previous collider experiments.
Excesses of events are observed at the low end of the explored mass range, around masses of 119 and
126\GeV, and at high mass around 320\GeV. These excesses, although not statistically significant,
make the observed limits weaker than expected in the absence of a signal.
At low mass, only the region $114.4 < \mH < 134\GeV$ remains consistent
with the expectation for the standard model Higgs boson production.

We wish to congratulate our colleagues in the CERN accelerator departments for the excellent performance of the LHC machine.
We thank the technical and administrative staff at CERN and other CMS institutes,
and acknowledge support from:
FMSR (Austria); FNRS and FWO (Belgium); CNPq, CAPES, FAPERJ, and FAPESP (Brazil);
MES (Bulgaria); CERN; CAS, MoST, and NSFC (China); COLCIENCIAS (Colombia); MSES (Croatia);
RPF (Cyprus); Academy of Sciences and NICPB (Estonia); Academy of Finland, MEC, and HIP (Finland);
CEA and CNRS/IN2P3 (France); BMBF, DFG, and HGF (Germany); GSRT (Greece); OTKA and NKTH (Hungary);
DAE and DST (India); IPM (Iran); SFI (Ireland); INFN (Italy); NRF and WCU (Korea); LAS (Lithuania);
CINVESTAV, CONACYT, SEP, and UASLP-FAI (Mexico); MSI (New Zealand); PAEC (Pakistan);
MSHE and NSC (Poland); FCT (Portugal); JINR (Armenia, Belarus, Georgia, Ukraine, Uzbekistan);
MON, RosAtom, RAS and RFBR (Russia); MSTD (Serbia); MICINN and CPAN (Spain); Swiss Funding Agencies (Switzerland);
NSC (Taipei); TUBITAK and TAEK (Turkey); STFC (United Kingdom); DOE and NSF (USA).

\bibliography{auto_generated}   % will be created by the tdr script.

\cleardoublepage \appendix\section{The CMS Collaboration \label{app:collab}}\begin{sloppypar}\hyphenpenalty=5000\widowpenalty=500\clubpenalty=5000\textbf{Yerevan Physics Institute,  Yerevan,  Armenia}\\*[0pt]
S.~Chatrchyan, V.~Khachatryan, A.M.~Sirunyan, A.~Tumasyan
\vskip\cmsinstskip
\textbf{Institut f\"{u}r Hochenergiephysik der OeAW,  Wien,  Austria}\\*[0pt]
W.~Adam, T.~Bergauer, M.~Dragicevic, J.~Er\"{o}, C.~Fabjan, M.~Friedl, R.~Fr\"{u}hwirth, V.M.~Ghete, J.~Hammer\cmsAuthorMark{1}, M.~Hoch, N.~H\"{o}rmann, J.~Hrubec, M.~Jeitler, W.~Kiesenhofer, M.~Krammer, D.~Liko, I.~Mikulec, M.~Pernicka$^{\textrm{\dag}}$, B.~Rahbaran, C.~Rohringer, H.~Rohringer, R.~Sch\"{o}fbeck, J.~Strauss, A.~Taurok, F.~Teischinger, P.~Wagner, W.~Waltenberger, G.~Walzel, E.~Widl, C.-E.~Wulz
% ...
\vskip\cmsinstskip
\dag:~Deceased\\
1:~~Also at CERN, European Organization for Nuclear Research, Geneva, Switzerland\\
% ...

\end{sloppypar}
\end{document}